\documentclass[showpacs,preprintnumbers,amsmath,amssymb,preprint]{revtex4}
\usepackage{amsfonts}
\usepackage{amssymb}
\usepackage{graphicx}
\usepackage[cmtip,arrow]{xy}
\newlength{\vshift}
\newlength{\hshift}
\usepackage{amssymb,amsopn}

\begin{document}
\title{Deformation Quantization of Fermi Fields}

%\author{ I. Galaviz$^1$}\email{igalaviz@fis.cinvestav.mx} \affiliation{Departamento de F\'{\i}sica,
%Centro de Investigaci\'on y de Estudios Avanzados del IPN\\
%P.O. Box 14-740, 07000 M\'exico D.F., M\'exico}

\author{I. Galaviz$^2$, H. Garc\'{\i}a-Compe\'an$^{1,2}$}
\email{igalaviz@fis.cinvestav.mx, compean@fis.cinvestav.mx}
\affiliation{$^1$Centro de Investigaci\'on y de Estudios Avanzados
del
IPN, Unidad Monterrey\\
Cerro de las Mitras 2565, cp. 64060, Col. Obispado, Monterrey
N.L., M\'exico}
\affiliation{$^2$Departamento de F\'{\i}sica,
Centro de Investigaci\'on y de Estudios Avanzados del IPN\\
P.O. Box 14-740, 07000 M\'exico D.F., M\'exico}

\author{M. Przanowski}\email{przan@fis.cinvestav.mx}
\affiliation{Institute of Physics\\
Technical University of \L \'od\'z\\
W\'olcza\'nska 219, 93-005, \L \'od\'z, Poland}

\author{F.J. Turrubiates}\email{fturrub@esfm.ipn.mx}
\affiliation{Departamento de F\'{\i}sica, Escuela Superior de
F\'{\i}sica y Matem\'aticas
del I.P.N.\\
Unidad Adolfo L\'opez Mateos, Edificio 9, 07738, M\'exico D.F.,
M\'exico\\}
%\date{\today}

\begin{abstract}
Deformation quantization for any Grassmann scalar free field is
described via the Weyl-Wigner-Moyal formalism. The
Stratonovich-Weyl quantizer, the Moyal $\star$-product and the
Wigner functional are obtained by extending the formalism proposed
recently in \cite{imelda} to the fermionic systems of infinite
number of degrees of freedom. In particular, this formalism is
applied to quantize the Dirac free field. It is observed that the
use of suitable oscillator variables facilitates considerably the
procedure. The Stratonovich-Weyl quantizer, the Moyal
$\star$-product, the Wigner functional, the normal ordering
operator, and finally, the Dirac propagator have been found with
the use of these variables.
\end{abstract}
\vskip -1truecm 

\maketitle

\newpage
\setcounter{equation}{0}

%%%%%%%%%%%%%%%%%%%%%%%%%%%%%%%%%%%%%%%%%%%%%%%%%%%%%%%%%%%%%%%%%%%%%%%%%%%%%%%
%%%%%%%%%%%%%%%%%%%%%%%%%%%%%%%%%%%%%%%%%%%%%%%%%%%%%%%%%%%%%%%%%%%%%%%%%%%%%%%
\section{Introduction}

Deformation quantization is commonly regarded as an alternative
approach to quantization. In principle, it can be implemented into
any classical system of particles, fields, strings or string
fields. This is based on the philosophy of deforming suitable
standard mathematical structures. Such a philosophy was introduced
in 1978 by Bayen, Flato, Fronsdal, Lichnerowicz and Sternheimer
\cite{bffls} (for a recent account, see \cite{disq,reviews}). The
deformation quantization formalism has a firm mathematical basis.
However, its application to quantize an arbitrary physical system
has still great challenges (see the third reference from
\cite{reviews}).

The canonical realization of deformation quantization is done with the use
of the Weyl-Wigner-Moyal (WWM) formalism introduced
originally in Refs. \cite{wwmoriginal}, which establishes a
one-one correspondence between the operator algebra (on certain
Hilbert space) and the algebra of symbols of operators
via the so called WWM correspondence. The product of operators is
mapped into an associative and noncommutative product of the respective
symbols called the $\star$-product. The simplest example is here
the famous Moyal $\star$-product.
Quantum theory in terms of deformation quantization has been intensively
studied for physical systems with a finite number of degrees of freedom.

Recently, much more interesting and difficult case of classical
fields (with an infinite number of degrees of freedom) has been
also considered by some authors. For example, deformation
quantization of scalar field is done in Refs.
\cite{dito,ditothree,ditodos,cfz,zachosone}. In some of these
references it has been shown that the $\star$-product can be
modified to absorb the divergences of the field theory with the
$\lambda \phi^4$-interaction. This modification is accomplished by
a $\lambda$ dependent redefinition of the $\star$-product
consistent with a cocycle condition. Free electromagnetic field in
deformation quantization within the Coulomb gauge has been studied
in \cite{campos}. Deformation quantization of gravitational field
as a constrained system has been discussed in
\cite{antonu,antond}. Then the linearized gravitational field from
the point of view of deformation quantization has been described
in \cite{hquevedo}. It is worth while to note that also
perturbation quantum field theory can be given the form of
deformation quantization formalism \cite{paftadq,ditofour,pqft}.
Moreover, as has been shown in \cite{strings} deformation
quantization theory can be applied to quantize the classical
bosonic strings in the light-cone gauge. On the other hand, string
theory uses also the deformation quantization formalism
(WWM-formalism) to describe the spacetime worldvolume of a
D-brane. The presence of a non-zero constant $B$-field on the
worldvolume deforms the product of functions (or classical fields)
on the D-brane into a Moyal $\star$-product, and the ordinary
effective field theory on the D-brane turns into a noncommutative
field theory with this Moyal $\star$-product \cite{sw}.

Most of the cases studied by the deformation quantization for
systems with a finite number of degrees of freedom deal with
bosonic variables. However, the analysis of some classical
physical systems requires the description of fermionic degrees of
freedom which involve Grassmann variables. These systems have been
discussed in the literature for several years
\cite{berezinbook,casa,marinov,sberezin,dewittbook,susyqmb}. (Some
extension of the WWM formalism to infinite degrees of freedom for
fermions can also be found in \cite{bereone,beretwo,schmutz}). The
canonical quantization of the fermionic systems by using the
Grassmann variables has been studied in all detail in
\cite{marnelius}. Very recently some authors started to apply the
machinery of deformation quantization to quantize the classical
fermionic systems
\cite{bordemann,bordemannII,hirshfeld,duetsch,zachosfermions,clifford}.
In particular, in \cite{hirshfeld,imelda} it has been shown
concretely how the deformation quantization program can be carried
over to specific physical fermionic systems. The same techniques
have been applied recently for the noncommutative superspace
\cite{seiberg}.

In the present paper we study the canonical approach to
deformation quantization for fermionic fields by employing the
traditional Weyl-Wigner-Moyal correspondence
\cite{wwmformalism,tata,hillery}. To this end we extend the
results obtained in the previous paper \cite{imelda} in the
context of fermionic systems with a finite number of degrees of
freedom. We deal with the Dirac free field and reproduce several
results about its quantization and its Green's functions.

Our paper is organized as follows. Sec. \ref{scalar_field} is
devoted to deformation quantization of a generic Fermi field.
First, we construct the Stratonovich-Weyl (SW) quantizer for this
Fermi field. Using a modified notion of the {\it trace} of an
operator one finds that the properties of the fermionic SW
quantizer have a similar form as the corresponding properties in
the bosonic case. Then the Moyal $\star$-product and the Wigner
functional are found. Finally, the normal ordering for the generic
Fermi field is briefly discussed. In Sec. \ref{dirac} the
deformation quantization formalism developed in Sec.
\ref{scalar_field} is applied to the case of the Dirac free field.
Using the oscillator variables we were able to simplify the
considerations. In particular, with the use of these variables we
have found the Wigner functional corresponding to the ground state
and also the Wigner functional of an arbitrary excited state.
Finally, in this section the Dirac propagator within the
deformation quantization is computed. Some conclusions and final
remarks in Sec. \ref{remarks} close the paper.

%%%%%%%%%%%%%%%%%%%%%%%%%%%%%%%%%%%%%%%%%%%%%%%%%%%%%%%%%%%%%%%
%%%%%%%%%%%%%%%%%%%%%%%%%%%%%%%%%%%%%%%%%%%%%%%%%%%%%%%%%%%%%%%
\vskip 2truecm
\section{Deformation Quantization of Grassmann Scalar
Field}\label{scalar_field}

Consider a scalar Grassmann field on the Minkowski spacetime
$M^{d+1}$ of signature $(-,-,\dots,-,+).$ By a Grassmann scalar
field we will understand a smooth function $\Theta$ over $M^{d+1}$
which takes values in the {\it field} of (anti-commuting)
Grassmann numbers $\mathbb{G}$, {\it i.e.}, $\Theta$ is the map
$\Theta: M^{d+1} \to \mathbb{G}$. Canonical variables of this
classical Grassmann field will be denoted by $\Theta(\vec{x},t)$
and $\pi_{\Theta} (\vec{x},t)$ with $(\vec{x},t) \in M^{d+1}=
\mathbb{R}^d \times \mathbb{R}.$

We deal with fields at the instant $t=0$ and we denote
$\Theta(\vec{x},0) \equiv \Theta(\vec{x})$ and $\pi_{\Theta}
(\vec{x},0) \equiv \pi_{\Theta}(\vec{x})$. It is worth while to
mention that some of the functional formulas and their
manipulations are formal. It is also important to notice that
since we will deal with Grassmann variables all the computations
and results obtained in the present paper are valid under the
specified conventions and ordering of factors given in this
section. In the present section we study the deformation
quantization of the Grassmann fields, including: the
Stratonovich-Weyl quantizer, the Moyal $\star$-product, the Wigner
functional \cite{wwmformalism,tata,hillery}, and the normal
ordering. In this paper we follow the conventions and notation
used in Ref. \cite{imelda}.

\vskip 1truecm
%%%%%%%%%%%%%%%%%%%%
\subsection{The Stratonovich-Weyl Quantizer} Let
$F[\pi_{\Theta},\Theta]$ be a functional on the phase space ${\cal
Z}_{\mathbb{G}}\equiv \{ (\pi_{\Theta}, \Theta)\}$ and let
$\widetilde{F}[\lambda, \mu]$ be its Fourier transform
\begin{equation}
\widetilde{F}[\lambda,\mu] = \int F[\pi_{\Theta}, \Theta ]
\exp\bigg\{-i \int d^d x \bigg( \pi_{\Theta}(\vec{x}) \cdot
\lambda(\vec{x}) +
 \Theta(\vec{x}) \cdot \mu(\vec{x}) \bigg) \bigg\} \prod d \pi_ {\Theta }  d \Theta,
\label{uno}
\end{equation}
where $\lambda(\vec{x})$ and $\mu(\vec{x})$ are the Fourier
transformed Grassmann fields corresponding to
$\pi_{\Theta}(\vec{x})$ and $\Theta(\vec{x})$ respectively,
$\pi_\Theta(\vec{x}) \cdot \lambda(\vec{x}) = \sum_{\alpha =1}^N
\pi_{\Theta \alpha}(\vec{x}) \lambda_\alpha(\vec{x})$ and the
functional measure $\prod d \pi_ {\Theta }  d \Theta :=
\prod_{\vec{x},\alpha} d \pi_{\Theta \alpha}(\vec{x}) d
\Theta_\alpha(\vec{x}).$  By the analogy to quantum mechanics we
can define the Weyl quantization rule as follows
\begin{equation}
\widehat{F}= W(F[\pi_{\Theta},\Theta]) := \int
\widetilde{F}[\lambda,\mu] \  \exp\bigg\{i\int d^d x \bigg(
\widehat{\pi}_{\Theta}(\vec{x}) \cdot \lambda(\vec{x}) +
\widehat{\Theta}(\vec{x}) \cdot \mu(\vec{x}) \bigg)\bigg \} \prod
d{\lambda}  d \mu,  \label{tres}
\end{equation}
with $\widehat{\pi}_{\Theta}$ and $\widehat{\Theta}$ being the
field operators  given by $\widehat{\pi}_{\Theta}(\vec{x})
|\pi_{\Theta}\rangle = \pi_{\Theta}(\vec{x}) |\pi_{\Theta}
\rangle$ and $\widehat{\Theta}(\vec{x}) |\Theta\rangle =
\Theta(\vec{x}) |\Theta \rangle$.

The coherent state $|\Theta \rangle$ can be defined in terms of
the vacuum state $|0\rangle$ \cite{wein,imelda}
\begin{equation}
|\Theta\rangle = \exp \left\{-{i\over\hbar} \int d^d x
\widehat{\pi}_\Theta (\vec{x}) \cdot \Theta (\vec{x}) \right\} | 0
\rangle, \label{defini}
\end{equation}
with $|\Theta\rangle\in {\cal F}_F$, where ${\cal F}_F= {\cal H}_F
\oplus {\cal H}_F\otimes {\cal H}_F \oplus \cdots$ is the Fock
space and ${\cal H}_F$ is a Hilbert space of fermions. By using
Eq. (\ref{defini}) it is easy to see that the functional state
satisfies the following property
\begin{equation}
\exp \left\{-{i\over\hbar}\int d^d x \widehat{\pi}_\Theta
(\vec{x}) \xi (\vec{x}) \right\} | \Theta \rangle = |\Theta + \xi
\rangle. \label{transla}
\end{equation}

It is also known that $(|\Theta \rangle)^* \neq \langle \Theta|$,
\cite{wein,imelda}. The dual Fock space ${\cal F}^*_F$ is
constructed from the dual vacuum state $\langle 0|$ as follows
\begin{eqnarray}
\langle \Theta | &=& \langle 0 | \prod_{\vec{x},\alpha}
\widehat{\Theta}_\alpha (\vec{x}) \exp \left\{-{i\over\hbar} \int
d^d x \Theta
(\vec{x}) \cdot \widehat{\pi}_\Theta (\vec{x}) \right\}\nonumber \\
&=& \langle 0 | \prod_{\vec{x},\alpha} \widehat{\Theta}_\alpha
(\vec{x}) \exp \left\{{i\over\hbar} \int d^d x
\widehat{\pi}_\Theta (\vec{x}) \cdot \Theta (\vec{x}) \right\}.
\label{bra}
\end{eqnarray}
This bra satisfies $\langle \Theta | \widehat{\Theta}_\alpha
(\vec{x}) = \langle \Theta | \Theta_\alpha (\vec{x})$ and
\begin{equation}
\langle \Theta' | \Theta \rangle = \prod_{\vec{x},\alpha}
\bigg(\Theta_\alpha (\vec{x}) - \Theta'_\alpha (\vec{x})\bigg) =:
\delta (\Theta - \Theta').
\end{equation}
Analogously for the momentum representation we define
\begin{equation}
|\pi_\Theta \rangle = \exp \left\{ -{i\over\hbar} \int d^d x
\widehat{\Theta} (\vec{x}) \cdot \pi_\Theta (\vec{x}) \right\}
\prod_{\vec{x},\alpha} \widehat{\pi}_{\Theta_\alpha} (\vec{x}) |0
\rangle, \ \ \ \ \ \ \langle \pi_\Theta| = \langle 0 | \exp
\left\{ {i\over\hbar} \int d^d x \widehat{\Theta} (\vec{x}) \cdot
\pi_\Theta (\vec{x}) \right\}.
\end{equation}
Then one gets
\begin{equation}
\langle \pi'_\Theta | \pi_\Theta \rangle = \prod_{\vec{x},\alpha}
\bigg(\pi'_{\Theta_\alpha} (\vec{x}) - \pi_{\Theta_\alpha}
(\vec{x})\bigg) =: \delta (\pi'_{\Theta} - \pi_\Theta).
\end{equation}
From the above relations it is easy to compute
\begin{equation}
\langle \pi_\Theta | \Theta \rangle = \exp \left\{ -{i\over\hbar}
\int d^d x  \pi_\Theta (\vec{x}) \cdot \Theta (\vec{x}) \right\},
\hspace{1.0cm} \langle \Theta | \pi_\Theta \rangle =
(i^\infty\hbar)^\infty\exp \left\{ {i\over\hbar} \int d^d x
\pi_\Theta (\vec{x}) \cdot \Theta (\vec{x}) \right\}.
\label{relat}
\end{equation}
With the use of (\ref{relat}) one can obtain the following completeness
relations
\begin{equation}
\int  |\pi_{\Theta} \rangle (-1)^\infty {\cal D} \pi_{\Theta}
\langle \pi_{\Theta} | = \widehat{1}, \hspace{2.0cm} \int | \Theta
\rangle {\cal D} \Theta \langle \Theta | = \widehat{1},
\label{complete}
\end{equation}
where ${\cal D}\Theta := \widetilde{\prod}_{\vec{x},\alpha} d
\Theta_\alpha(\vec{x}),$ ${\cal D} \pi_{\Theta} :=
\widetilde{\prod}_{\vec{x},\alpha} d\pi_{\Theta \alpha} (\vec{x})$
and $\widetilde{\prod}_{\vec{x},\alpha}$ means that the
differentials are ordered oppositely to the ordering of the
Grassmann variables in the integrated function.

Return to our description of the Weyl correspondence.
Substituting (\ref{uno}) into (\ref{tres}) one gets
\begin{equation}
\widehat{F} = W(F[\pi_{\Theta},\Theta])= \int
F[\pi_{\Theta},\Theta] \widehat{\Omega} [\pi_{\Theta},
\Theta] \prod d\pi_ {\Theta} d\Theta ,
 \label{trece}
\end{equation}
where $\widehat{\Omega}$ is the Stratonovich-Weyl quantizer (see
Refs. \cite{imelda,wwmformalism})
\begin{equation}
\widehat{\Omega} [\pi_{\Theta},\Theta] = \int \exp \bigg \{ -i
\int d^d x \bigg( \pi_{\Theta}(\vec{x})\cdot \lambda (\vec{x}) +
{\Theta} (\vec{x})\cdot \mu (\vec{x}) \bigg) \bigg \}
\exp\bigg\{i\int d^d x \bigg( \widehat{\pi}_{\Theta}(\vec{x})
\cdot \lambda(\vec{x}) + \widehat{\Theta}(\vec{x})\cdot
\mu(\vec{x}) \bigg)\bigg\}\prod d\lambda d\mu .
\label{catorce}
\end{equation}

Making use of the well known Campbell-Baker-Hausdorff formula for
Grassmann variables in the appropriate ordering, the commutation
rules for $\widehat{\pi}_{\Theta}$ and $\widehat{\Theta}$ and the
relations (\ref{transla}) and (\ref{complete}), it is possible to
reexpress $\widehat{\Omega}[\pi_{\Theta},\Theta]$ in a very useful
form
$$
\widehat{\Omega}[\pi_{\Theta},\Theta] = (-i)^\infty \int {\cal D}
\mu \exp \bigg \{ -i \int d^d x \ \Theta(\vec{x}) \cdot
\mu(\vec{x}) \bigg \} \bigg{|} \pi_{\Theta} - {\hbar\mu \over 2}
\bigg\rangle \bigg\langle \pi_{\Theta} + {\hbar\mu \over
2}\bigg{|}
$$
\begin{equation}
= (i)^\infty  \int {\cal D} \lambda \exp \bigg \{ - i \int d^d x
\pi_{\Theta}(\vec{x}) \cdot \lambda(\vec{x}) \bigg\} \bigg{|}
\Theta - {\hbar \lambda \over 2} \bigg\rangle \bigg\langle \Theta
+ {\hbar \lambda \over 2} \bigg{|}. \label{dnueve}
\end{equation}
Now let us define the "trace" as follows
\begin{eqnarray}
{\rm tr}\{ \widehat{\cal O}\} &:=& (i\hbar)^{-\infty}\int{\cal D}
\Theta
\big\langle \Theta \big| \widehat{\cal O} \big| \Theta \big\rangle\nonumber\\
&=&(i\hbar)^{-\infty}\int{\cal D} \pi_{\Theta} \big\langle
\pi_{\Theta} \big| \widehat{\cal O} \big| \pi_{\Theta}
\big\rangle, \nonumber
\end{eqnarray}
for any operator $\widehat{\cal O}$ (compare with \cite {imelda}).

With this definition one can check also that the
Stratonovich-Weyl operator satisfies the following properties

\begin{equation}
{\rm tr} \big \{ \widehat{\Omega}[\pi_{\Theta},\Theta] \big \} =
1, \label{dseis}
\end{equation}
\begin{equation}
{\rm tr} \bigg \{ \widehat{\Omega}[\pi_{\Theta},\Theta]
\widehat{\Omega}[\pi_{\Theta} ', \Theta '] \bigg \} =
\prod_{\vec{x},\alpha} \big(\Theta_\alpha(\vec{x}) -
\Theta'_\alpha(\vec{x})\big) \big(\pi_{{\Theta}_\alpha}(\vec{x}) -
\pi'_{{\Theta}_\alpha}(\vec{x})\big) =: \delta (\Theta - \Theta' ,
\pi_{\Theta} - \pi_{\Theta} '). \label{dsiete}
\end{equation}

Finally, multiplying (\ref{trece}) by
$\widehat{\Omega}[\pi_{\Theta},\Theta]$ and taking into account
the property (\ref{dsiete}) one easily gets the fundamental relation
\begin{equation}
F[\pi_{\Theta},\Theta] = {\rm tr} \bigg \{ \widehat{\Omega}
[\pi_{\Theta},\Theta] \widehat{F} \bigg\}. \label{docho}
\end{equation}

\vskip .5truecm
%%%%%%%%%%%%%%%%%%%%%%%%%%%%%
\subsection{The Moyal $\star$-Product}\label{SP}

We are at the position to define the Moyal $\star$-product in
field theory involving Grassmann scalar fields. Let
$F=F[\pi_{\Theta},\Theta]$ and $G=G[\pi_{\Theta},\Theta]$ be some
functionals on $\cal{Z}_{\mathbb{G}}$ that correspond to the field
operators $\widehat{F}$ and $\widehat{G}$ respectively, i.e.
$F[\pi_{\Theta},\Theta]=W^{-1}(\widehat{F})={\rm tr} \bigg \{
\widehat{\Omega} [\pi_{\Theta},\Theta] \widehat{F} \bigg \}$ and
$G[\pi_{\Theta},\Theta]=W^{-1}(\widehat{G}) = {\rm tr} \bigg \{
\widehat{\Omega}[\pi_{\Theta},\Theta] \widehat{G} \bigg \}$. The
functional which corresponds to the operator product $\widehat{F}
\widehat{G}$ will be denoted by $(F \star
G)[\pi_{\Theta},\Theta]$. So we have
\begin{equation}
(F \star G)[\pi_{\Theta},\Theta]:= W^{-1}(\widehat{F}
\widehat{G})= {\rm tr} \bigg \{
\widehat{\Omega}[\pi_{\Theta},\Theta] \widehat{F} \widehat{G}
\bigg \}. \label{vsiete}
\end{equation}
Using Eqs. (\ref{trece})  and (\ref{vsiete})  and performing some
simple but rather lengthy calculations one gets
$$
(F \star G) [\pi_{\Theta},\Theta] = \bigg({ i\hbar \over 2}
\bigg)^{2\infty}\int \ F[\pi_{\Theta} ', \Theta '] G[\pi_{\Theta}
'', \Theta ''] $$
\begin{equation}
\times \exp \bigg \{ -{2i \over \hbar} \int d^d x \bigg(
\pi_{\Theta} (\Theta' - \Theta'' ) + \pi_{\Theta}' (\Theta'' -
\Theta ) + \pi_{\Theta}''(\Theta - \Theta ') \bigg) \bigg \} \prod
d\pi_{\Theta} ' d\Theta ' \prod d\pi_{\Theta}'' d \Theta '' .
\label{vocho}
\end{equation}

Introduce new variables: $\Psi ' = \Theta ' -
\Theta,$ $\Psi '' = \Theta '' - \Theta,$  $\Pi ' = \pi_{\Theta} '
- \pi_{\Theta}$, $\Pi ''= \pi_{\Theta}'' - \pi_{\Theta} .$ Using
the expansion of $F[\pi_{\Theta} ', \Theta '] = F[\pi_{\Theta} +
\Pi ', \Theta + \Psi ']$ and $G[\pi_{\Theta} '', \Theta ''] =
G[\pi_{\Theta} + \Pi '', \Theta + \Psi ''] $ into the Taylor series and
performing some manipulations we obtain
\begin{equation}
\big(F \star  G\big)[\pi_{\Theta},\Theta] = F[\pi_{\Theta},\Theta]
\exp\bigg\{{i\hbar\over 2} \buildrel{\leftrightarrow}\over {\cal
P}_{\mathbb{G}}\bigg\} G[\pi_{\Theta},\Theta], \label{treinta}
\end{equation}
with
\begin{equation}
\buildrel{\leftrightarrow}\over {\cal P}_{\mathbb{G}} :=  \int d^d
x \bigg({{\buildrel{\leftarrow}\over {\delta}}\over \delta
\Theta(\vec{x})} {{\buildrel{\rightarrow}\over {\delta}}\over
\delta \pi_{\Theta}(\vec{x})} + {{\buildrel{\leftarrow}\over
{\delta}}\over \delta \pi_{\Theta}(\vec{x})}
{{\buildrel{\rightarrow}\over {\delta}}\over \delta
\Theta(\vec{x})}\bigg), \label{tuno}
\end{equation}
where $\overleftarrow{\delta}$ and $\overrightarrow{\delta}$ are
the right and left  functional derivatives, respectively. The last
formula corresponds exactly to the Poisson bracket for two
functionals $F$ and $G$ which is given by
$$
F\buildrel{\leftrightarrow}\over {\cal P}_{\mathbb{G}} G :=\{
F[\pi_{\Theta},\Theta],G[\pi_{\Theta},\Theta] \}_{P}
$$
\begin{equation}
= (-1)^{\epsilon_{F}} \int d^d x \bigg\{ {\delta
F[\pi_{\Theta},\Theta] \over \delta\Theta(\vec{x})}{\delta
G[\pi_{\Theta},\Theta] \over \delta \pi_{\Theta}(\vec{x})}
 + {\delta
F[\pi_{\Theta},\Theta] \over \delta \pi_{\Theta}(\vec{x})} {\delta
G[\pi_{\Theta},\Theta] \over \delta \Theta(\vec{x})}\bigg\},
\label{poissonsusy}
\end{equation}
with $\epsilon_F=0,1$ depending on whether the corresponding functional
$F$ is even or odd, respectively. This Poisson bracket has been
reported in the literature for many years
\cite{casa,marinov,sberezin,dewittbook,bordemann,bordemannII,duetsch,zachosfermions,
hirshfeld,clifford,imelda}.

The associated symplectic form
\begin{equation}
\omega_{\mathbb{G}} = \int d^d x\  \delta \Theta(\vec{x}) \wedge
\delta \pi_{\Theta}(\vec{x}), \label{symstruct}
\end{equation}
provides ${\cal Z}_{\mathbb{G}}$ with the structure of a
symplectic supermanifold.

\vskip .5truecm
%%%%%%%%%%%%%%%%%%%%%%%%%%
\subsection{The Wigner Functional}

If $\widehat{\rho}$ is the density operator of a quantum state
then the functional  $\rho_{_W}[\pi_{\Theta},\Theta]$ defined by
\begin{equation}
\rho_{_W}[\pi_{\Theta},\Theta] = {\rm tr} \bigg \{
\widehat{\Omega}[\pi_{\Theta},\Theta] \widehat{\rho}\bigg \} =
2^{-\infty} \int {\cal D} {\eta}
\exp \bigg\{ - {2i \over \hbar} \int d^d x \ \pi_{\Theta}(\vec{x})
\cdot \eta(\vec{x}) \bigg\} \langle \Theta +\eta| \widehat{\rho} | \Theta - \eta
\rangle
\label{ttres}
\end{equation}
is called the Wigner functional corresponding to this state.
In particular, the Wigner functional $\rho_{_W}[\pi_{\Theta},\Theta]$
corresponding to the quantum state $\widehat{\rho} = |\Phi\rangle
\langle \Phi^\dag|$ , where $\langle \Phi^\dag| := (|\Phi\rangle)^\dag $ , is given by
\cite{wwmoriginal,tata,hillery,cfz,imelda}
\begin{equation}
\rho_{_W}[\pi_{\Theta},\Theta] =2^{-\infty} \int {\cal D} {\eta}
\exp \bigg\{ - {2i \over \hbar} \int d^d x \ \pi_{\Theta}(\vec{x})
\cdot \eta(\vec{x}) \bigg\} \Phi[\Theta + \eta ] \Phi^\dag[\Theta
- \eta ], \label{tcinco}
\end{equation}
where $ \Phi[\Theta + \eta ] = \langle \Theta +\eta| \Phi \rangle$
and $\Phi^\dag[\Theta - \eta ] = \langle \Phi^\dag | \Theta - \eta
\rangle$.

\vskip .5truecm
%%%%%%%%%%%%%%%%%%%%%%%%%%
\subsection{Normal Ordering}
Let $F[\pi_{\Theta},\Theta]$ be a functional on ${\cal
Z}_\mathbb{G}$. We define
\begin{equation}
F_{\widehat{\cal N}} := \widehat{\cal N}F,
\end{equation}
The operator $\widehat{\cal N}$ is the relativistic field generalization
of the respective operator given in \cite{imelda} by the formula (64), and in the
present case it reads
\begin{equation}
\widehat{\cal N} := \exp \left\{ {1\over 2} \sum_{i=1}^N \int d^d
p f(E_{\vec{p}}) \bigg( {\overrightarrow{\delta}^2 \over \delta
b_i(\vec{p}) \delta b_i^*(\vec{p})} - {\overrightarrow{\delta}^2
\over \delta d_i^*(\vec{p}) \delta d_i(\vec{p})} \bigg) \right\},
\end{equation}
where $f(E_{\vec{p}})$ is a function to be determined,
$b_i(\vec{p})$, $b_i^*(\vec{p})$ stand for the $W^{-1}-images$ of
the annihilation or creation operators (respectively) for
particles, and $ d_i(\vec{p})$, $d_i^*(\vec{p})$ denote the
$W^{-1}-images $ of annihilation or creation operators for
antiparticles.

Then the Berezin-Wick or normal quantization is given by
$$
\widehat{F}_{\widehat{\cal N}} = \int F_{\widehat{\cal N}}
\widehat\Omega[\pi_\Theta,\Theta] \prod d\pi_{\Theta} d\Theta,
$$
and the Weyl mapping of $F_{\widehat{\cal N}}$ gives the normal
ordering of the Weyl image of $F_{\widehat{\cal N}}$ i.e.
\begin{equation}
: \widehat{F} :=: W(F):\ \buildrel{df}\over {=} W(F_{\widehat
{\cal N}}) \ \buildrel{df}\over {=} W_{\widehat {\cal N}} (F).
\end{equation}
In the next section the normal ordering in the case of the free
Dirac field will be considered in more detail.

\vskip 1truecm
%%%%%%%%%%%%%%%%%%%%%%%%%%%%%%%%%%%%%%%%%%%%%%%%%%%%%%%%%%%%%%%%%%%%%
%%%%%%%%%%%%%%%%%%%%%%%%%%%%%%%%%%%%%%%%%%%%%%%%%%%%%%%%%%%%%%%%%%%%%
\section{Deformation Quantization of the Dirac Free Field}\label{dirac}

The aim of this section is to provide an example of the
application of the deformation quantization of Grassmann fields to
the Dirac free field. In addition we compute the propagator of the
Dirac field in this context.

In the next subsection we briefly survey the Dirac field in order
to introduce the notation and conventions that we use in this
paper. We stress the uses of the oscillator variables ${\bf b}^*$
and ${\bf b}$ which allowed us to perform the construction
\cite{hatfield}.

%%%%%%%%%%%%%%%%%%%%%%%%%%%%%%%%%%%%%%%%%%%%%%%
\subsection{The Dirac Free Field}

In this section we discuss the Dirac free field $\psi(x)$ over
Minkowski spacetime $M^{3+1}=\mathbb{R}^3 \times \mathbb{R}$ with
the signature $(-,-,-,+)$ and $x=(\vec{x},t) \in M$. The action is
given by
$$
I_D[\psi] = \int d^3xdt {\cal L}_D$$
\begin{equation}
 = \int d^3xdt \overline{\psi}(\vec{x},t)\cdot \hbar c \big(i
\not \!
\partial -{mc\over \hbar} \big) \psi(\vec{x},t),
\label{diraclagran}
\end{equation}
where $\not \! \partial=\gamma^{\mu}\partial_{\mu}$,
$\gamma^{\mu}$ are the Dirac matrices ($\mu = 0, \dots, 3$),
$\overline{\psi}({x}) \equiv \psi^{\dag}({x}) \gamma^0$, $m$ is
the mass parameter, $c$ is the speed of light  and $\hbar$ the
Planck constant. Thus, the field $\psi({x})$ fulfills the Dirac
equation
\begin{equation}
\big(i\not \! \partial - {mc\over\hbar} \big) \psi (\vec{x},t) =
0. \label{diraceqn}
\end{equation}
Its {\it conjugate momentum} is given by $\pi_{\psi}(\vec{x},t) =
{-\partial {\cal L}_D \over \partial (
\partial \dot{\psi}) } = i \hbar {\psi}^{\dag}(\vec{x},t)$, where $
\dot{\psi}(\vec{x},t) \equiv {\partial \psi(\vec{x},t) \over
\partial t}$ \footnote{The conjugate momentum $\widetilde{\pi}_{\psi}$
of the field $\psi$ is defined by \cite{hirshfeld,imelda}
$$
\widetilde{\pi}_{\psi_{\alpha}} = {\partial{\cal L} \over
\partial(\partial_t \psi_{\alpha})} = -i \hbar \psi_{\alpha}^\dag,
$$
where the sign $(-)$ appears as one consider $\overline{\psi}$ and
$\psi$ as Grassmann variables. In field theory is usual to write
down it as $\pi = i \hbar \psi^\dag$, see for instance,
\cite{wein,hatfield}. Therefore to be consistent with these
references  on quantum fields we choice $ \psi \mathrm{\ \ and\ \
} \pi_{\psi}= i \hbar \psi^\dag,$ as our fundamental variables,
instead of $\psi$ and $\widetilde{\pi}_{\psi}$.}. Then the
Hamiltonian reads
$$
H_D[\pi_{\psi}, \psi] = \int d^3x \ \overline{\psi}(\vec{x},t)
\cdot \hbar c \big(-i\gamma^j\partial_j + {mc\over\hbar}
\big)\psi(\vec{x},t)
$$
$$
 = \int d^3x \ \psi^{\dag}(\vec{x},t) \cdot i \hbar
{\partial \over
\partial t} \psi(\vec{x},t)
$$
\begin{equation}
= \int d^3x \ \pi_{\psi}(\vec{x},t) \cdot {\partial \over
\partial t} \psi(\vec{x},t),
\label{dhamilton}
\end{equation}
where as usual $ \pi_\psi \cdot \partial_t \psi \equiv
\sum_{\alpha=1}^4 \pi_{\psi_\alpha} \partial_t \psi_\alpha, $ and
$\gamma^i = \beta \alpha^i$, $\gamma^0 = \beta,$ with $i=1,2,3$.
Note that we use the Weyl (or chiral) representation of the Dirac
matrices $\gamma^\mu,$ $\mu=0,1,2,3$
\begin{equation}
\begin{array}{ccc}
  \gamma^0 = \left(%
\begin{array}{cc}
  1 & 0 \\
  0 & -1 \\
\end{array}%
\right), & \hspace{0.5cm}  \gamma^j = \left(%
\begin{array}{cc}
  0 & \sigma_j \\
  -\sigma_j & 0 \\
\end{array}%
\right), &\hspace{0.5cm} j=1,2,3 \\
\end{array}
\end{equation}
with $\sigma_j$ being Pauli's matrices. The Dirac matrices have
the following properties: $ {\gamma^j}^\dag = -\gamma^j,$
${\gamma^0}^\dag=\gamma^0,$ $ \gamma^\mu \gamma^\nu + \gamma^\nu
\gamma^\mu = 2\eta^{\mu\nu}.$
%%%%%%%%%%%%%%%

According to the definition of the Poisson bracket for Grassmann
fields given by (\ref{poissonsusy}), the Poisson bracket
corresponding to the Dirac free field takes the form
$$
\{F,G\}_P = F \ {\buildrel \leftrightarrow \over {\cal P}}\ G
$$
where
\begin{eqnarray}
{\buildrel \leftrightarrow \over {\cal P}} &=& \int d^3x \bigg(
{{\buildrel \leftarrow \over \delta}\over \delta \psi(\vec{x},t)}
{{\buildrel \rightarrow \over \delta}\over \delta
\pi_{\psi}(\vec{x},t)} + {{\buildrel \leftarrow \over \delta}\over
\delta \pi_{\psi}(\vec{x},t)} {{\buildrel
\rightarrow \over \delta}\over \delta \psi(\vec{x},t)} \bigg)\nonumber\\
&=&-{i \over \hbar} \int d^3x \bigg( {{\buildrel \leftarrow \over
\delta}\over \delta \psi(\vec{x},t)} {{\buildrel \rightarrow \over
\delta}\over \delta \psi^\dag(\vec{x},t)} + {{\buildrel \leftarrow
\over \delta}\over \delta \psi^\dag(\vec{x},t)} {{\buildrel
\rightarrow \over \delta}\over \delta \psi(\vec{x},t)} \bigg),
\label{PB}
\end{eqnarray}
and ${{\buildrel \leftarrow \over \delta}\over \delta
\psi(\vec{x},t)} {{\buildrel \rightarrow \over \delta}\over \delta
\pi_{\psi}(\vec{x},t)} \equiv \sum_{\alpha =1}^4 {{\buildrel
\leftarrow \over \delta}\over \delta \psi_{\alpha}(\vec{x},t)}
{{\buildrel \rightarrow \over \delta}\over \delta
\pi_{\psi_{\alpha}}(\vec{x},t)},$ etc.

Consequently, for $\psi_{\alpha}(\vec{x},t)$ and
${\pi_{\psi}}_{\alpha} (\vec{x},t)$ one gets
$$
\{ \psi_{\alpha} (\vec{x},t), {\pi_{\psi}}_{\beta} (\vec{y},t)
\}_{P} =  \delta (\vec{x}-\vec{y}) \delta_{\alpha\beta},
$$
\begin{equation}
\{ \psi_{\alpha} (\vec{x},t), \psi_{\beta} (\vec{y},t) \}_{P} = 0,
\hspace{1.0cm} \{
{\pi_{\psi}}_{\beta}(\vec{x},t),{\pi_{\psi}}_{\beta}(\vec{y},t)
\}_{P}=0. \label{comm}
\end{equation}
(Remember that $\alpha=1,2,3,4$ runs over the components of the
Dirac spinor).

Usually quantization is done by the substitution $
{\{\cdot,\cdot\}}_P \mapsto {1\over i\hbar} {[\cdot,\cdot]}_+$.
Thus we obtain
\begin{equation}
\begin{array}{l}
{[\widehat{\psi}_\alpha(\vec{x},t),
\widehat{\psi}_\beta(\vec{y},t)]}_+ =
{[\widehat{\pi}_{\psi_\alpha}(\vec{x},t),
\widehat{\pi}_{\psi_\beta}
(\vec{y},t)]}_+ = 0, \\
{[\widehat{\psi}_\alpha(\vec{x},t), \widehat{\pi}_{\psi_\beta}
(\vec{y},t)]}_+
= i\hbar \delta_{\alpha\beta} \delta(\vec{x}-\vec{y}),\\
{[\widehat{\psi}_\alpha(\vec{x},t),
\widehat{\psi}_\beta^\dag(\vec{y},t)]}_+ =
 \delta_{\alpha\beta} \delta(\vec{x}-\vec{y}).\\
\end{array}
\end{equation}

The field variables $\psi_{\alpha}(\vec{x},t)$ and
${\pi_{\psi}}_{\alpha} (\vec{x},t)$ can be expanded into the plane
waves
\begin{equation}
 \psi_{\alpha}(\vec{x},t) = \sum_{i=1}^2 \int {d^3p
\over (2 \pi\hbar)^{3}} {m c^2\over E_{\vec{p}}} \bigg[ {b}_i
(\vec{p},t) u_{\alpha}^i(\vec{p}) e ^{i\vec{p}\cdot\vec{x}/ \hbar
} + {d}_i^* (\vec{p},t) {v_{\alpha}^i}(\vec{p})
e^{-i\vec{p}\cdot\vec{x}/\hbar} \bigg], \label{fieldone}
\end{equation}
\begin{equation}
 {\pi_{\psi}}_{\alpha} (\vec{x},t) = i \hbar
{\psi}_{\alpha}^{\dag}(\vec{x},t) = \sum_{i=1}^2 \int {d^3p \over
(2 \pi\hbar)^{3}} \ i \hbar {mc^2 \over E_{\vec{p}}}
\bigg[{b}_i^{*}(\vec{p},t) {{u}_{\alpha}^i}^{\dag}(\vec{p})
e^{-i\vec{p}\cdot\vec{x}/\hbar } + {d}_i(\vec{p},t)
{{v}_{\alpha}^i}^{\dag}(\vec{p}) e^{i\vec{p}\cdot\vec{x}/ \hbar
}\bigg], \label{fieldtwo}
\end{equation}
where $b_i(\vec{p},t)=b_i(\vec{p})\exp{(-iE_{\vec{p}}t/\hbar)},$
$d_i^*(\vec{p},t)=d_i^*(\vec{p})\exp{(iE_{\vec{p}}t/\hbar)}$
and $E_{\vec{p}} =\sqrt{c^2\vec{p}^2 + m^2c^4}$. Here the index
$i=1,2$ stands for the spin degrees of freedom. Therefore, $u^i$ and
$v^i$ are solutions of the Dirac equation of momentum $p$
with positive or negative energies, respectively
\begin{equation}
(\not \! \! p-mc)u^i=0,\ \ \ \ \ \ \ \ \ \ (\not \! \!
p+mc)v^i=0, \label{pnspinor}
\end{equation}
with $\not \! \! p= \gamma^{\mu}p_{\mu}$; and they are restricted
to satisfy the following relations
$$
{u^i}^{\dag}(\vec{p})u^j(\vec{p})=
{v^i}^{\dag}(\vec{p})v^j(\vec{p})={E_{\vec{p}}
\over mc^2}\delta_{ij},
$$
\begin{equation}
\overline{v}^i(\vec{p})u^j(\vec{p})=0.
\label{relations}
\end{equation}
Classically Dirac fields are described by Grassmann variables
since they are functions of the anti-commuting Grassmann variables
$b_i(\vec{p}), b_i^*(\vec{p}) ,d_i(\vec{p})$ and $d_i^*(\vec{p})$.
Poisson brackets for these variables are

\begin{equation}
\{b_i(\vec{p}),b_j^*(\vec{p}\ ') \}_P =
\{d_i(\vec{p}),d_j^*(\vec{p}\ ')\}_P= -{i \over \hbar} (2
\pi\hbar)^3\bigg({E_{\vec{p}}\over mc^2}
\bigg)\delta_{ij}\delta(\vec{p}-\vec{p}\ '). \label{grass}
\end{equation}

\vskip .5truecm \noindent {\it Oscillator Variables}

The variables $b$ and $d$ are quite asymmetric and it is not easy to
express these variables in terms of the field variables $\psi$ and
$\pi_{\psi}$. To carry out the deformation quantization in the
simplest possible way we introduce the variables ${\bf
b}(\vec{p},t,r)$ and ${\bf b}^{*}(\vec{p},t,r)$ \cite{hatfield}
which are related to the variables $b$ and $d$ as follows
$${\bf b}(\vec{p},t,r)= (2 \pi\hbar)^{-3/2}\sqrt{ mc^2
\over E_{\vec{p}}} b_r(\vec{p},t), \hspace{1.0cm} {\rm with\ }
r=1,2,$$ and
$${\bf b}(\vec{p},t,r)= (2 \pi\hbar)^{-3/2}\sqrt{ mc^2 \over
E_{\vec{p}}} d^*_{r-2}(-\vec{p},t), \hspace{1.0cm} {\rm with\ }
r=3,4;$$ and ${\bf b}^*$ is determined by the complex conjugate of
these equations.

These variables suggest to redefine the solutions to the Dirac
equation in the following form
$$
w_{\alpha}(\vec{p},r)={u_{\alpha}}^r(\vec{p}), \ \ \  {\rm for } \
\ r=1,2,
$$
\begin{equation}
 w_{\alpha}(\vec{p},r)= v_{\alpha}^{r-2}(-\vec{p}), \
\ \ {\rm for } \ \ r=3,4. \label{orelat}
\end{equation}
From (\ref{relations}) one gets the relations
$$
{w}^{\dag}(\vec{p},r) w(\vec{p},r')=
{E_{\vec{p}} \over mc^2} \delta_{rr'} ,
$$
$$
\bar{w}(\varepsilon_r \vec{p},r)
w(\varepsilon_r' \vec{p},r')=\varepsilon_r \delta_{rr'},
$$
\begin{equation}
\sum_{r=1}^4 w_{\alpha}(\vec{p},r) w_{\alpha'}^{\dag}(\vec{p},r)=
{E_{\vec{p}}\over mc^2 }\delta_{\alpha \alpha'}. \label{ident}
\end{equation}

In terms of these variables Eqs. (\ref{fieldone}) and
(\ref{fieldtwo}) read
\begin{equation} \psi_{\alpha}(\vec{x},t) = \sum_{r=1}^4 \int
{d^3p \over (2 \pi\hbar)^{3/2}} \sqrt{mc^2 \over E_{\vec{p}}} {\bf
b}(\vec{p},t,r)w_{\alpha}(\vec{p},r) \exp \big( i
\vec{p}\cdot\vec{x}/\hbar \big), \label{fpone}
\end{equation}
\begin{equation}
 {\pi_{\psi}}_{\alpha} (\vec{x},t) = i \hbar
{\psi}_{\alpha}^{\dag}(\vec{x},t) = \sum_{r=1}^4 \int {d^3p \over
(2 \pi\hbar)^{3/2}} \ i \hbar \sqrt{mc^2 \over E_{\vec{p}}} {\bf
b}^{*}(\vec{p},t,r) {w}_{\alpha}^{\dag}(\vec{p},r)
\exp\big(-i\vec{p}\cdot\vec{x}/\hbar \big), \label{fptwo}
\end{equation}
where ${\bf b}(\vec{p},t,r) = {\bf b}(\vec{p},r)\exp \big\{ -i
\varepsilon_r E_{\vec{p}}\ t /\hbar\big\}$ with $\varepsilon_r=1$
for $r=1,2$ and $\varepsilon_r=-1$ for $r=3,4$.

Substituting Eqs. (\ref{fpone}) and (\ref{fptwo}) into (\ref{comm})
we find the Poisson brackets for the variables ${\bf b}$ y ${\bf
b}^{*}$
$$
\{ {\bf b}(\vec{p},r), {\bf b}^{*}(\vec{p}\ ',r') \}_{P} = -{i
\over \hbar} \delta (\vec{p} - \vec{p}\ ') \delta_{rr'},
$$
\begin{equation}
 \{ {\bf b}(\vec{p},r), {\bf b}(\vec{p}\ ',r')
\}_{P} = 0, \ \ \ \ \ \{ {\bf b}^{*}(\vec{p},r), {\bf
b}^{*}(\vec{p}\ ',r') \}_{P}= 0. \label{ocommrela}
\end{equation}
Multiplying Eq. (\ref{fpone}) by $w_{\beta}^{\dag}(\vec{p}\ ',r')
\exp \big(-i\vec{p}\ '\cdot\vec{x}/\hbar\big)$ and integrating
over $\mathbb{R}^3$ we get
$$
\int d^3x \ w_{\beta}^{\dag}(\vec{p\
}',r')\psi_{\alpha}(\vec{x},t) \exp \big( -i\vec{p}\
'\cdot\vec{x}/\hbar \big)
$$
$$
= \sum_{r=1}^4 \int {d^3x \ d^3p \over (2 \pi\hbar)^{3/2}}
\sqrt{mc^2 \over E_{\vec{p}}} w_{\beta}^{\dag} (\vec{p}\ ',r')
{\bf b}(\vec{p},t,r)w_{\alpha}(\vec{p},r) \exp \big(
i(\vec{p}-\vec{p}\ ')\cdot\vec{x}/\hbar \big). \label{comput}
$$
Using now the relations (\ref{ident}), after some computations
we obtain
\begin{equation}
 {\bf b}(\vec{p},t,r)=\sqrt{m c^2 \over E_{\vec{p}}}
\sum_{\alpha}\int {d^3x \over (2\pi\hbar)^{3/2}}\
w_{\alpha}^{\dag}(\vec{p},r) \psi_{\alpha}(\vec{x},t) \exp
\big(-i\vec{p}\cdot\vec{x}/\hbar \big), \label{annihi}
\end{equation}
\begin{equation}
 {\bf b}^{*}(\vec{p},t,r)=\sqrt{mc^2 \over E_{\vec{p}}}
\sum_{\alpha} \int {d^3x \over (2\pi\hbar)^{3/2}}\
\psi_{\alpha}^{\dag}(\vec{x},t) w_{\alpha}(\vec{p},r)
 \exp \big(i\vec{p}\cdot\vec{x}/\hbar \big).
\label{creat}
\end{equation}
Thus the Grassmann variables ${\bf b}$ and ${\bf b}^{*}$,
determine precisely the canonical conjugate variables which we
will use to describe the Dirac field in the WWM formalism.

Substituting (\ref{fpone}) and (\ref{fptwo}) into the Hamiltonian
(\ref{dhamilton}) yields
\begin{equation}
 H_D[{\bf b}^*,{\bf
b}]=  \sum_{r=1}^4 \int {d^3p} \varepsilon_r E_{\vec{p}} {\bf
b}^*(\vec{p},r) {\bf b}(\vec{p},r), \label{hamione}
\end{equation}
or in terms of the standard variables,
\begin{equation}
H_D[b,b^*,d,d^*]= \sum_{i=1}^2 \int {d^3p \over (2\pi\hbar)^{3}}
mc^2 \bigg[ {b}_i^*(\vec{p}) {b}_i(\vec{p}) - {d}_i(\vec{p})
{d}_i^*(\vec{p}) \bigg]. \label{hamitwo}
\end{equation}

\vskip .5truecm
%%%%%%%%%%%%%%%%%%%%%%%%%%%%%%%%%%%%%%%%%%%%%%%%
%%%%%%%%%%%%%%%%%%%%%%%%%%%%%%%%%%%%%%%%%%%%%%%
\subsection{The Stratonovich-Weyl Quantizer}

The Weyl-Wigner-Moyal correspondence for the Dirac field in the
field variables reads
\begin{equation}
\widehat{F} = W(F[\pi_{\psi},\psi])= W(F[{\psi}^{\dag},\psi])=
\big({i\hbar}\big)^{-\infty} \int F[{\psi}^{\dag},\psi]
\widehat{\Omega} [{\psi}^{\dag},\psi] \prod d{\psi}^{\dag} d\psi ,
\label{operatorone}
\end{equation}
where
$$
\widehat{\Omega}[{\psi}^{\dag},\psi] = (-i)^\infty\int {\cal D}
\mu \exp \bigg \{ -{i } \int d^3x \psi(\vec{x}) \cdot \mu(\vec{x})
\bigg \} \bigg|i \hbar {\psi}^{\dag} - {\hbar\mu \over 2}
\bigg\rangle \bigg\langle i \hbar {\psi}^{\dag} + {\hbar\mu \over
2}\bigg|
$$
\begin{equation}
 = (i)^\infty \int {\cal D} \lambda \exp \bigg \{\hbar \int d^3x
 {\psi}^{\dag}(\vec{x}) \cdot \lambda(\vec{x})  \bigg \} \bigg|\psi
 -{\hbar \lambda \over 2} \bigg\rangle \bigg\langle \psi + {\hbar \lambda \over 2}\bigg|.
\label{stratono}
\end{equation}
In terms of the oscillator variables we have
\begin{equation}
\widehat{F} = W(F[{\bf b}^*,{\bf b}])=
\big({i\hbar}\big)^{-\infty} \int \prod {d\bf b}^* {d\bf b} F[{\bf
b}^*,{\bf b}] \widehat{\Omega} [{\bf b}^*,{\bf b}],
\label{operatorsb}
\end{equation}
and
$$
\widehat{\Omega}[{\bf b}^*,{\bf b}] =(-i)^\infty \int {\cal D}
\chi  \exp \bigg \{ -i\sum_{r=1}^4 \int {d^3p} {\bf b}(\vec{p},r)
\chi(\vec{p},r) \bigg \} \bigg|i \hbar{\bf b}^* - {\hbar\chi \over
2} \bigg\rangle \bigg\langle  i \hbar {\bf b}^* + {\hbar\chi \over
2}\bigg|
$$
\begin{equation}
 = (i)^\infty \int {\cal D} \xi \exp \bigg \{ \hbar \sum_{r=1}^4\int
 {d^3p}{\bf b}^*(\vec{p},r) \xi(\vec{p},r) \bigg
\} \bigg|{\bf b} - {\hbar\xi \over 2} \bigg\rangle \bigg\langle
{\bf b} + {\hbar\xi \over 2}\bigg|, \label{stratobs}
\end{equation}
where $\chi$ and  $\xi$ are Dirac spinors.

\noindent
[Observe that here $\infty$ is in fact $4\infty$ as
$r=4$. Consequently, one can put 1 instead of $i^\infty$ or
$(-i)^\infty$].

\vskip .5truecm
%%%%%%%%%%%%%%%%%%%%%%%%%%%%%%%%%%%%%%%%%%%%%%%
\subsection{The Moyal $\star$-Product}

According to the WWM correspondence, the symbol of an operator
$\widehat{F}$ is given by
\begin{equation}
F[{\bf b}^*,{\bf b} ] = W^{-1}(\widehat{F}) = {\rm tr}
\bigg\{\widehat{\Omega}[{\bf b}^*,{\bf b}] \widehat{F} \bigg\}.
\label{invmap}
\end{equation}
The Moyal $\star$-product in this case can be defined similarly as
it has been done in (\ref{treinta}). Let $F_1[{\bf b}^*,{\bf b}]$
and $F_2[{\bf b}^*,{\bf b}]$ be functionals over the Dirac phase
space defined by: ${\cal Z}_D = \{(
{\pi_{\psi}}_{\alpha}(\vec{x}),\psi_{\alpha}(\vec{x}))_{\vec{x}\in
\mathbb{R}^3}\} = \{(i \hbar {\bf b}^*(\vec{p},r), ({\bf
b}(\vec{p},r))_{r=1,\cdots,4}\}$ and let $\widehat{F}_1$ and
$\widehat{F}_2$ be their corresponding operators. Then by a
similar computation to that done in Sec. {\ref{SP}} we finally get
\begin{equation}
\big(F_1 \star F_2\big)[{\bf b}^*,{\bf b}] = F_1[{\bf b}^*,{\bf
b}] \exp\bigg({i\hbar\over 2} \buildrel{\leftrightarrow} \over
{\cal P}_D\bigg) F_2[{\bf b}^*,{\bf b}], \label{fmoyalp}
\end{equation}
where
\begin{equation}
\buildrel{\leftrightarrow}\over {\cal P}_D := -{i \over \hbar}
\sum_{r=1}^{4}\int {d^3p }\bigg( {{\buildrel{\leftarrow}\over
{\delta}}\over \delta {\bf b}(\vec{p},r)}
{{\buildrel{\rightarrow}\over {\delta}}\over \delta {\bf
b}^*(\vec{p},r)} + {{\buildrel{\leftarrow}\over {\delta}}\over
\delta {\bf b}^*(\vec{p},r)} {{\buildrel{\rightarrow}\over
{\delta}}\over \delta {\bf b}(\vec{p},r)}  \bigg),
\label{dpoisson}
\end{equation}
is the Poisson operator. In
terms of the standard Grassmann variables $b$, $b^*$, $d$ and
$d^*$, it can be rewritten as
\begin{equation}
\buildrel{\leftrightarrow}\over {\cal P}_D = - {i\over \hbar}
(2\pi\hbar)^3 \sum_{i=1}^{2}\int {d^3p } \bigg( {E_{\vec{p}} \over
mc^2} \bigg) \bigg \{ \bigg( {{\buildrel{\leftarrow}\over
{\delta}}\over \delta {b}_i(\vec{p})}
{{\buildrel{\rightarrow}\over {\delta}}\over \delta
{b}_i^*(\vec{p})} + {{\buildrel{\leftarrow}\over {\delta}}\over
\delta {b}_i^*(\vec{p})} {{\buildrel{\rightarrow}\over
{\delta}}\over \delta {b}_i(\vec{p})} \bigg)
 +  \bigg( {{\buildrel{\leftarrow}\over {\delta}}\over
\delta {d}_i^*(\vec{p})} {{\buildrel{\rightarrow}\over
{\delta}}\over \delta {d}_i(\vec{p})} +
{{\buildrel{\leftarrow}\over {\delta}}\over \delta {d}_i(\vec{p})}
{{\buildrel{\rightarrow}\over {\delta}}\over \delta
{d}_i^*(\vec{p})} \bigg)\bigg\}. \label{dpoissonbs}
\end{equation}
The operator $\buildrel{\leftrightarrow}\over {\cal P}_D$
determines the Poisson bracket
$\{F,G\}_P=F\buildrel{\leftrightarrow} \over {\cal P}_D G$ defined
by the symplectic structure
$$
\omega_D=\int d^3x \sum_{\alpha} \delta \psi_{\alpha}(\vec{x})
\wedge \delta \pi_{\psi}(\vec{x})
$$
\begin{equation}
= i \hbar \int d^3x \sum_{\alpha} \delta \psi_{\alpha}(\vec{x})
\wedge \delta \psi_{\alpha}^{\dag}(\vec{x}). \label{dsymplstru}
\end{equation}
This symplectic structure defines the symplectic supermanifold
$({\cal Z}_D,\omega_D)$ which represents the phase space. In terms of
the original field variables $\psi_{\alpha}^{\dag},\psi_{\alpha}$
one has
\begin{equation}
\buildrel{\leftrightarrow}\over {\cal P}_D := - {i \over \hbar}
\sum_{\alpha}\int d^3x \bigg({{\buildrel{\leftarrow}\over
{\delta}}\over \delta \psi_{\alpha}(\vec{x})}
{{\buildrel{\rightarrow}\over {\delta}}\over \delta
\psi_{\alpha}^{\dag}(\vec{x})} + {{{\buildrel{\leftarrow}\over
{\delta}}\over \delta \psi_{\alpha}^{\dag}(\vec{x})}
{{\buildrel{\rightarrow}\over {\delta}}\over \delta
\psi_{\alpha}(\vec{x})}}\bigg). \label{anotherpoiss}
\end{equation}
\vskip .5truecm
%%%%%%%%%%%%%%%%%%%%%%%%%%%%%%%%%%%%%%%%%%%%%%%
\subsection{The Wigner Functional}

The Wigner functional for the Dirac free field is defined in
analogy to the scalar field case. Let $\widehat{\rho}^{phys}$ be
the density operator corresponding to the quantum physical state
of the Dirac field. Then the Wigner functional corresponding to
this state is given by
$$
\rho_{_W}[{\bf b}^*,{\bf b}] = {\rm tr} \bigg\{
\widehat{\Omega}[{\bf b}^*,{\bf b}] \widehat{\rho}^{phys}\bigg\}
$$
\begin{equation}
 = 2^{-\infty} \int {\cal D} \xi \exp
\bigg\{ -{2}\sum_{r=1}^4 \int d^3p \ {\bf b}^*(\vec{p},r)
\xi(\vec{p},r) \bigg\} \langle {\bf b} - \xi |
\widehat{\rho}^{phys}| {\bf b} + \xi \rangle. \label{wignerf}
\end{equation}
In the case when $\widehat{\rho}^{phys} = |\Phi \rangle \langle
\Phi^\dag|$ the above equation gives
\begin{equation}
\rho_{_W}[{\bf b}^*,{\bf b}] = 2^{-\infty} \int {\cal D} \xi \exp
\bigg\{-{2} \sum_{r=1}^4 \int d^3p \ {\bf b}^*(\vec{p},r)
\xi(\vec{p},r)  \bigg\} \Phi[{\bf b} - \xi] \Phi^{\dag}[{\bf b} +
\xi], \label{wignerftwo}
\end{equation}
where $\Phi[{\bf b}]=\langle {\bf b} | \Phi \rangle$ and
$\Phi^\dag[{\bf b}]=\langle  \Phi^\dag |{\bf b}  \rangle$.

\noindent [Note that since $r$ runs through $1,2,3$ and $4$ i.e.,
through an even number of indices the Jacobian of transformation
leading from (\ref{ttres}) or (\ref{tcinco}) to (\ref{wignerf}) or
(\ref{wignerftwo}), respectively, is equal to $1$.]

\vskip 1truecm
%%%%%%%%%%%%%%%%%%%%%%%%%%%%%%%%%%%%%%%%%%%%%%%%
\noindent {\it Example: The Ground State}

In the Schr\"odinger representation the ground state corresponding
to the system described by the Dirac equation can be easily found.
In order to do that we use the wave functional defined by
$\Phi[\psi] = \langle \psi | \Phi \rangle$.Then the canonical
conjugate momentum to $\psi$ is represented by $i \hbar
\widehat{\psi}^{\dag}= i \hbar{\delta \over \delta \psi}$.
Substituting this momentum into the Hamiltonian (\ref{dhamilton})
we get the time independent Schr\"odinger equation
\begin{equation}
\int d^3x \left(\hbar {\delta \over \delta \psi(\vec{x})}c(-i{\bf
{\buildrel \rightarrow \over \alpha}  \cdot {\buildrel \rightarrow
\over \nabla} } + \beta {mc\over\hbar}) \psi(\vec{x})\right)
\Phi[\psi]= E\Phi[\psi]. \label{fdiraceq}
\end{equation}

It is an easy matter to show that Eqs. (\ref{fpone}) and
(\ref{fptwo}) at $t=0$ can be rewritten in terms of respective
operators in the Schr\"odinger representation as follows
$$
\widehat{\psi}(\vec{x}) = \sum_{r=1}^4 \int {d^3p \over (2
\pi\hbar)^{3/2}} \sqrt{mc^2 \over E_{\vec{p}}}\widehat{\bf
b}(\vec{p},r)w(\vec{p},r) \exp \big(i\vec{p} \cdot \vec{x} / \hbar
\big),
$$
$$
\widehat{\psi}^{\dag}(\vec{x})= {\delta \over \delta
\psi(\vec{x})} = \sum_{r=1}^4 \int {d^3p \over (2 \pi\hbar)^{3/2}}
\ \sqrt{mc^2 \over E_{\vec{p}}} {\delta \over \delta {\bf
b}(\vec{p},r) } w^{\dag}(\vec{p},r) \exp\big(-i \vec{p} \cdot
\vec{x} /\hbar \big).
$$
Inserting these formulae into (\ref{fdiraceq}) one gets
\begin{equation}
\sum_{r=1}^4 \int d^3p \varepsilon_r E_{\vec{p}} {\delta \over
\delta {\bf b}(\vec{p},r)} {\bf b}(\vec{p},r)\Phi[{\bf b}] = E
\Phi[{\bf b}], \label{functdeq}
\end{equation}
where ${\bf b}(\vec{p},r)$ is the eigenvalue of $\widehat{\bf
b}(\vec{p},r)$.

The stable vacuum $|0 \rangle$ for the Dirac free field is the
ground state in which all the negative energy states are filled.
To be precise consider the case of having only one fermion state
and remind that ${\bf b}$ and $\hbar {\delta \over \delta {\bf
b}}$ are the annihilation and creation operators of a single
fermion. In addition, let $\Omega_0({\bf b})$ and $\Omega_1({\bf
b})$ be the (one fermion) wave functions of the states $|0\rangle$
and $|1\rangle$, respectively. Hence one has
$$
{\bf b}\Omega_0({\bf b})=0,  \ \ \ \ \ \ \ \ \ {\delta \over
\delta {\bf b}}\Omega_1({\bf b})=0,
$$
\begin{equation}
{\bf b}\Omega_1({\bf b})=\Omega_0({\bf b}),  \ \ \ \ \ \ \ \
{\delta \over \delta {\bf b}} \Omega_0({\bf b}) =\Omega_1({\bf
b}). \label{groundone}
\end{equation}

Since ${\bf b}$ is a Grassmann variable it is easy to see that
$\Omega_0({\bf b})={\bf b}$ and $\Omega_1({\bf b})=1$. In order to
be consistent with the structure of the stable vacuum the
positive energy states ($r=1,2$) are determined by the function
${\bf b}$, and therefore they are empty. While for the negative
energy states ($r=3,4$), the function is $1$, and consequently
they are occupied.

The Hamiltonian is a sum over all momenta $p$'s and
states $r$'s. This implies that the ground state is given by an
infinite product
\begin{equation}
\Phi_0[{\bf b}]=\eta \prod_{r=1}^2\prod_{\vec{p}} {\bf
b}(\vec{p},r), \label{infprod}
\end{equation}
where $\eta$ is a normalization factor. Thus the adjoint
functional is given by
\begin{equation}
\Phi_0^{*}[{\bf b}]=\eta^*\prod_{r=3}^4\prod_{\vec{p}} {\bf
b}(\vec{p},r). \label{adjprod}
\end{equation}

Now we are at the position to compute the form of the Wigner
functional of the ground state corresponding to the Dirac field.
Substituting the wave functionals (\ref{infprod}) and
(\ref{adjprod}) into (\ref{wignerftwo})and taking $|\eta|^2 =1$
one finally gets
$$
{\rho_{W}}_{0}[{\bf b}^*,{\bf b}]  = 2^{-\infty} \int {\cal D} \xi
\exp \bigg\{ -2 \sum_{r=1}^4 \int {d^3p } \ {\bf b}^*(\vec{p},r)
\xi(\vec{p},r)  \bigg\} \Phi_0[{\bf b} - \xi] \Phi^\dag_0[{\bf b}
+ \xi]
$$
\begin{equation}
= 2^{-\infty} \int {\cal D} \xi \exp \bigg\{ -2 \sum_{r=1}^4 \int
{d^3p } \ {\bf b}^*(\vec{p},r) \xi(\vec{p},r) \bigg\}
\prod_{r=1}^4 \prod_{\vec{p}}({\bf b}(\vec{p},r) - \varepsilon_r
\xi). \label{wigfgs}
\end{equation}
Recall that $\xi$ is also a Dirac spinor and ${\bf b}(\vec{p},r) +
\varepsilon_r \xi$ is proportional to the Dirac delta function.
Then the Wigner functional (including the normalization factor)
for the ground state is given by
\begin{equation}
{\rho_{W}}_{0}[{\bf b}^*,{\bf b}]  = 2^{-\infty} \exp \bigg\{- {2}
\int {d^3p } \sum_{r=1}^4 \varepsilon_r{\bf b}^*(\vec{p},r) {\bf
b}(\vec{p},r) \bigg\}. \label{wfgsb}
\end{equation}
In terms of the standard variables it yields
\begin{equation}
{\rho_{W}}_{0}[b,b^*,d,d^*]  = 2^{-\infty} \exp \bigg\{ -{2} \int
{d^3p \over (2\pi\hbar)^{3}}\bigg( {mc^2 \over E_{\vec{p}}} \bigg)
\sum_{i=1}^2\bigg( b_i^*(\vec{p}) b_i(\vec{p}) -
d_i(\vec{p})d_i^*(\vec{p}) \bigg)\bigg\}. \label{wfgsbd}
\end{equation}

Remark: One can quickly show that the ground state Wigner
functional ${\rho_{W}}_{0}[{\bf b}^*,{\bf b}]$ can also be found
within purely deformation quantization language from the following
equations
$$
{\bf b}(\vec{p},r=1,2)  \star {\rho_{W}}_{0}[{\bf b}^*,{\bf b}]= 0
\ \ \ \ {\rm and} \ \ \ \ {\bf b}^*(\vec{p},r=3,4) \star
{\rho_{W}}_{0}[{\bf b}^*,{\bf b}]= 0.
$$
\vskip 1truecm
%%%%%%%%%%%%%%%%%%%%%%%%%%%%%%%%%%%%%%%%%%%%%%%
\noindent {\it Wigner Functional for Excited States}

The Winger functionals for excited states can be constructed as
follows: If one wants to get the wave functional of an electron
(of positive energy) of momentum $\vec{p}$ and spin $s=\pm {1
\over 2}$, one have to remove from the ground state wave
functional $\Phi_0[{\bf b}]$ a corresponding factor ${\bf
b}(\vec{p},r=1,2)$, where $r=1$ corresponds to spin up and $r=2$
to spin down, and substitute it by the factor 1. Thus, the wave
functional for a state with one electron of momentum $\vec{p}$ and
spin up is: $\Phi_{1e}[{\bf b}]=\eta \prod_{r=1}^2\prod_{\vec{q}}
{\bf b}(\vec{q},r),$ except the factor ${\bf b}(\vec{p},1)$. If
you prefer to add a positron of momentum $\vec{p}$ and spin $s=\pm
{1 \over 2}$, it is necessary to annihilate an electron of
negative energy. In order to do that one needs to multiply
$\Phi_0[{\bf b}]$ by ${\bf b} (-\vec{p},r=4,3)$, where $r=4$
corresponds to the positron spin up and $r=3$ to spin down (for
details, see \cite{hatfield}). For example, the wave functional
for the positron of momentum $\vec{p}$ and spin up is given by:
$\Phi_{1p}[{\bf b}]=\eta {\bf b} (-\vec{p},r=4)
\prod_{r=1}^2\prod_{\vec{q}} {\bf b}(\vec{q},r)$. The Wigner
functionals corresponding to excited states can be obtained in
this context. Consider the higher quantum state of the Dirac field
describing  $n$ electrons of the momentum $p_{1},\dots,p_{n}$ with
some definite spin, and $n'$ positrons of the momentum
$p_{1'},\dots,p_{n'}$ and a definite spin: $|(\vec{p}_1,r_1),
\dots (\vec{p}_n,r_n), (\vec{p}_{1'},r_{1'}), \dots
(\vec{p}_{n'},r_{n'}) \rangle := \widehat{\bf
b}^{\dag}(\vec{p}_1,r_1) \cdots \widehat{\bf
b}^{\dag}(\vec{p}_n,r_n) \widehat{\bf b}(-\vec{p}_{1'},r_{1'})
\cdots
 \widehat{\bf b}(-\vec{p}_{n'},r_{n'}) |0\rangle$, where $r_{1},\dots,r_{n}$
take the values $1$ or $2$ and $r_{1'},\dots,r_{n'}$ take 3 or 4.
The respective density operator is given by
$$
\widehat{\rho} = \widehat{\bf b}^{\dag}(\vec{p}_1,r_1) \cdots
\widehat{\bf b}^{\dag}(\vec{p}_n,r_n) \widehat{\bf
b}(-\vec{p}_{1'},r_{1'}) \cdots
 \widehat{\bf b}(-\vec{p}_{n'},r_{n'})
$$
\begin{equation}
\times  |0\rangle \langle 0| \widehat{\bf
b}^{\dag}(-\vec{p}_{n'},r_{n'}) \cdots \widehat{\bf
b}^{\dag}(-\vec{p}_{1'},r_{1'}) \widehat{\bf b}(\vec{p}_n,r_n)
\cdots \widehat{\bf b}(\vec{p}_1,r_1).
\end{equation}

Thus the Wigner functional corresponding to our excited quantum
state reads
$$
{\rho_{W}}[{\bf b}^*,{\bf b}] ={\bf b}^{*}(\vec{p}_1,r_1)\star
\cdots \star {\bf b}^{*} (\vec{p}_n,r_n) \star {\bf
b}(-\vec{p}_{1'},r_{1'})\star \cdots \star {\bf
b}(-\vec{p}_{n'},r_{n'}) \star {\rho_{W}}_{0}
$$
\begin{equation}
\star{\bf b}^{*}(-\vec{p}_{n'},r_{n'}) \star \cdots \star {\bf
b}^{*}(-\vec{p}_{1'},r_{1'}) \star {\bf b}(\vec{p}_n,r_n)\star
\cdots \star {\bf b}(\vec{p}_1,r_1), \label{fexcststes}
\end{equation}
where ${\rho_{W}}_{0}$ is given by (\ref{wfgsb}). Due to the form of
the Moyal product (\ref{fmoyalp}) and (\ref{dpoisson})this Wigner functional
can be rewritten in the following form
$$
{\rho_{W}}[{\bf b}^*,{\bf b}] ={\bf b}^{*}(\vec{p}_1,r_1) \cdots
{\bf b}^{*} (\vec{p}_n,r_n) {\bf b}(-\vec{p}_{1'},r_{1'})\cdots
{\bf b}(-\vec{p}_{n'},r_{n'}) \star {\rho_{W}}_{0}
$$
\begin{equation}
\star {\bf b}^{*}(-\vec{p}_{n'},r_{n'})\cdots {\bf
b}^{*}(-\vec{p}_{1'},r_{1'})  {\bf b}(\vec{p}_n,r_n) \cdots {\bf
b}(\vec{p}_1,r_1).
\label{fexcststesII}
\end{equation}
From this last formula one can find the Wigner functional for any
excited state.

\vskip .5truecm

\noindent {\it Normal Ordering}

The normal ordering of the Dirac field operators can also be
defined in the deformation quantization formalism. This can be
done by using the operator $\widehat{\cal N}$ acting on the phase
space ${\cal Z}_D$
\begin{equation}
\widehat{\cal N} := \exp \bigg\{ {1\over 2} \int {d^3p}
\sum_{r=1}^4  {\varepsilon_r \stackrel{\to}{\delta}^2 \over \delta
{\bf b}(\vec{p},r) \delta {\bf b}^*(\vec{p},r)} \bigg \}.
\label{normalop}
\end{equation}
In terms of the standard variables
\begin{equation}
\widehat{\cal N} = \exp \bigg\{ {1\over 2}{(2\pi\hbar)^3} \int
{d^3p} \bigg( {E_{\vec{p}} \over mc^2} \bigg) \sum_{i=1}^2 \bigg(
{\stackrel{\to}{\delta}^2 \over \delta b_i(\vec{p}) \delta
b_i^*(\vec{p})} - {\stackrel{\to}{\delta}^2 \over \delta
d_i^*(\vec{p}) \delta d_i(\vec{p})}\bigg) \bigg \}.
\label{normalobds}
\end{equation}

Let $F[{\bf b}^*,{\bf b}]$ be a functional defined on the phase
space ${\cal Z}_D$. Then $F_{\widehat{\cal N}}$ is defined as follows
\begin{equation}
F_{\widehat{\cal N}}[{\bf b}^*,{\bf b}]=\widehat{\cal
N}F[{\bf b}^*,{\bf b}]. \label{sepone}
\end{equation}
Consequently, the Weyl image of $F_{\widehat{\cal N}}[{\bf b}^*,{\bf b}]$
gives the normal ordering of the Weyl image of $F[{\bf b}^*,{\bf b}]$
\begin{equation}
: \widehat{F} : = : W(F[{\bf b}^*,{\bf b}]):\ \buildrel{df}\over {=} W(F_{\widehat {\cal
N}}[{\bf b}^*,{\bf b}]) \ \buildrel{df}\over {=} W_{\widehat {\cal
N}} (F[{\bf b}^*,{\bf b}]). \label{septwo}
\end{equation}

\vskip 1.5truecm

\noindent
{\it Example: The Hamiltonian}

The normal ordering of the Hamiltonian can be obtained by applying (\ref{septwo})
with $\widehat{\cal N}$ given by Eq. (\ref{normalop}) to
$H_D[{\bf b}^*,{\bf b}]$ given by Eq. (\ref{hamione}). Simple calculations yield
\begin{equation}
H_{D_{\widehat{\cal N}}} = H_D +  2 \int {d^3p } E_{\vec{p}}
\delta(0). \label{sephamiltwo}
\end{equation}
Then the normal ordered Hamiltonian operator reads
\begin{equation}
:\widehat H_D:\ = \sum_{i=1}^2 \int {d^3p \over (2\pi \hbar)^3}
mc^2 \bigg(\widehat{b}^{\dag}_i(\vec{p}) \widehat{ b}_i(\vec{p}) +
\widehat{d}^{\dag}_i(\vec{p}) \widehat{d}_i(\vec{p}) \bigg).
\label{hamiltlast}
\end{equation}

\vskip 1truecm
%%%%%%%%%%%%%%%%%%%%%%%%%%%5
\subsection{Dirac Propagator}

In order to compute the propagator of the Dirac field we need to
find
\begin{equation}
 iS_F({x}-{y}) = \langle
0|\widehat{\psi}_{\alpha}({x}) \widehat{\overline{\psi}}_{\beta}({y})|0\rangle \cdot
\theta(t-t') - \langle 0|
\widehat{\overline{\psi}}_{\beta}({y}) \widehat{\psi}_{\alpha}({x})|0\rangle \cdot
\theta(t'-t). \label{propa}
\end{equation}
So we first compute the quantities $\langle 0|\widehat{\psi}_{\alpha}({x})
\widehat{\overline{\psi}}_{\beta}({y})|0\rangle $ and $\langle 0|
\widehat{\overline{\psi}}_{\beta}({y})\psi_{\alpha}({x})|0\rangle$. In terms
of deformation quantization these expectation values are given by (compare with
\cite{campos})

\begin{equation}
\langle 0 |\widehat{\psi}_{\alpha}({x}) \widehat{\overline{\psi}}_{\beta}({y})| 0
\rangle =
 {\int \prod d{\bf b}^* d {\bf b} \ \psi_{\alpha}({x})
 \star
\overline{\psi}_{\beta}({y}) \ \rho_{W_0} [{\bf b}^*,{\bf b}]
\over \int \prod d{\bf b}^* d {\bf b} \ \rho_{W_0} [{\bf b}^*,{\bf
b}]}, \label{promedio}
\end{equation}
and the analogous formula for the second expectation value.

Carrying out the corresponding integrations and making use of the
following relations
$$
\sum_{r=1}^2 w_{\alpha}(\vec{p},r) \overline{w}_{\beta}(\vec{p},r)
= {(\not \! \! p + mc)_{\alpha \beta} \over 2mc}, \hspace{1.5cm}
\sum_{r=3}^4 \overline{w}_{\alpha}(\vec{p},r)
{w}_{\beta}(\vec{p},r) = {(\not \! \! p - mc)_{\alpha \beta} \over
2mc}.
\label{secondfor}
$$
after straightforward calculations we arrive at the results which are
well known in quantum field theory
\begin{equation}
\langle 0 |\widehat{\psi}_{\alpha}({x})
\widehat{\overline{\psi}}_{\beta}({y})| 0 \rangle =  \int {d^3p
\over (2 \pi\hbar)^3}{c(\not \! \! p + mc)_{\alpha \beta} \over
2E_{\vec{p}}} \exp \big(-i p\cdot(x-y)/\hbar\big);
\label{cienvcinco}
\end{equation}
and
\begin{equation}
 \langle 0|
\widehat{\overline{\psi}}_{\beta}({y}) \widehat{\psi}_{\alpha}({x})|0\rangle =
  \int {d^3p \over (2 \pi\hbar)^3}{ c(\not \! \! p -
mc)_{\beta \alpha} \over 2E_{\vec{p}}} \exp \big(i
p\cdot(x-y)/\hbar\big). \label{otro}
\end{equation}
The above formulas reproduce exactly the propagator of the Dirac field in
the deformation quantization formalism.

%%%%%%%%%%%%%%%%%%%%%%%%%%%%%%%%%%%%%%%%%%%%%%%%%%%%%%%%%%%%%%%%%%%%%%%%%%

\section{Final Remarks}\label{remarks}

In this paper we have carried over the deformation quantization
program via the Weyl-Wigner-Moyal formalism to fermionic fields.
In a sense it is an extension of the results for a finite number
of degrees of freedom given in Ref. \cite{imelda} to field theory.
To see how our construction explicitly works we have applied it to
the quantization of the Dirac free field. The quantization of this
field is possible since the Dirac free field can be regarded as an
infinite number of decoupled Fermi oscillators. Consequently, the
prescription worked out in Refs. \cite{hirshfeld,imelda} can be
employed to each oscillator separately. The oscillator variables
${\bf b}$ and ${\bf b}^*$ greatly facilitates the procedure.

In the next step it is natural to consider a formalism which
involves both bosonic and fermionic fields, and also their
interactions in a similar way to the one given in Ref.
\cite{imelda} for the case of supersymmetric quantum mechanics.
Then, gathering the results of Ref. \cite{campos} concerning
deformation quantization of the electromagnetic field with those
of the present paper (in addition of some other considerations), a
version of QED in the context of deformation quantization can be
quickly implemented \cite{qed}.

We are aware that the quantization by deformation seems to be at
most equivalent to those of canonical or path integral
quantizations. However, there is some evidence that deformation
quantization would be more general. For instance, recently it was
found that one of the intriguing features of deformation
quantization is well defined for spaces with orbifold and conical
singularities \cite{fronsdal}. The results obtained in the present
paper would be important for the description of fermions on
noncommutative orbifolds \cite{martinec,belhaj}. This problem will
be reported elsewhere.

\vskip 2truecm
%%%%%%%%%%%%%%%%%%%%%%%%%%%%%%%%%%%%%%%%%%%%%%%%%%%%%%%%%%%%%%%%%%%%%%%%%%
\centerline{\bf Acknowledgments}

It is a pleasure to thank G. Dito and C. Maldonado-Mercado for
useful discussions. This work was supported in part by CONACyT
M\'exico Grant 45713-F.  I.G. wish to thank CINVESTAV, Unidad
Monterrey, for its hospitality where part of this work was done.
The research of I.G. is supported by a CONACyT graduate
fellowship.

%%%%%%%%%%%%%%%%%%%%%%%%%%%%%%%%%%%%%%%%%%%%%%%%%%%%%%%%%%%%%%%%%%%%%%%
%%%%%%%%%%%%%%%%%%%%%%%%%%%%%%%%%%%%%%%%%%%%%%%%%%%%%%%%%%%%%%%%%%%%%%%


\begin{references}

\bibitem{bffls} F. Bayen, M. Flato, C. Fronsdal, A. Lichnerowicz and D.
Sternheimer, {\it Ann. Phys.} {\bf 111}, 61 (1978); F. Bayen, M. Flato, C. Fronsdal, A. Lichnerowicz and D.
Sternheimer, {\it Ann. Phys.} {\bf 111}, 111 (1978).

\bibitem{disq} D. Sternheimer, ``Quantization is Deformation'',
talk
http://www.u-bourgogne.fr/monge/d.sternh/papers/DS\-ober\-wolfach.pdf.

\bibitem{reviews} C.K. Zachos, ``Deformation Quantization:
Quantum Mechanics Lives and Works in Phase Space'', Int. J. Mod.
Phys. A {\bf 17} (2002) 297, hep-th/0110114; A.C. Hirshfeld and P.
Henselder, ``Deformation Quantization in the Teaching for Quantum
Mechanics'', Am. J. Phys. {\bf 70} (2002) 537; G. Dito and D.
Sternheimer, ``Deformation Quantization: Genesis, Developments and
Metamorphoses, {\it Deformation Quantization} (Strasbourg 2001)
Lect. Math. Theor. Phys. {\bf 1} Ed. de Gruyter, Berlin, IRMA
(2002) pp. 9-54.

\bibitem{wwmoriginal} H. Weyl, {\it Group Theory and Quantum Mechanics}, (Dover,
New York, 1931); E.P. Wigner, {\it Phys. Rev.} {\bf 40}, 749
(1932); A. Groenewold, Physica {\bf 12} (1946) 405-460; J.E.
Moyal, {\it Proc. Camb. Phil. Soc.} {\bf 45}, 99 (1949).

\bibitem{dito} G. Dito, ``Star Product Approach to Quantum Field Theory:
The Free Scalar Field'', {\it Lett. Math. Phys.}
 {\bf 20}, 125 (1990).

\bibitem{ditothree} G. Dito, ``Star Products and Nonstandard Quantization
for Klein-Gordon Equation'', J. Math. Phys. {\bf 33} (1992) 791.

\bibitem{ditodos} G. Dito, ``An Example of Cancellation of Infinities in
the Star Quantization of Fields'', {\it Lett. Math. Phys.} {\bf
27}, 73 (1993).

\bibitem{cfz} T. Curtright D. Fairlie and C.K. Zachos, {\it Phys.
Rev.} {\bf D58},  025002 (1998).

\bibitem{zachosone} T. Curtright and C.K. Zachos, J. Phys.  A {\bf 32},
771 (1999).

\bibitem{campos} H. Garc\'{\i}a-Compe\'an, J.F. Pleba\'nski, M.
Przanowski and F.J. Turrubiates, ``Deformation Quantization of
Classical Fields'', Int. J. Mod. Phys. A {\bf 16} (2001) 2533.

\bibitem{antonu} F. Antonsen, {\it Phys. Rev.}  {\bf D56}, 920 (1997).

\bibitem{antond} F. Antonsen, ``Deformation Quantization of Constrained
Systems'', gr-qc/9710021.

\bibitem{hquevedo} H.~Quevedo and J.~G.~Tafoya, ``Towards the deformation quantization
of linearized gravity,'' Gen.\ Rel.\ Grav.\  {\bf 37}, 2083
(2005), [arXiv:gr-qc/0401088].

\bibitem{paftadq}M. Duetsch and K. Fredenhagen. ``Perturbative Algebraic
Field Theory, and Deformation Quantization'', hep-th/0101079.

\bibitem{ditofour} G. Dito, ``Deformation Quantization of Covariant Fields'',
math.qa/0202271.

\bibitem{pqft} A.C. Hirshfeld and P. Henselder, ``Star Products and
Perturbative Quantum Field Theory'',  Annals Phys. {\bf 298}
(2002) 382, hep-th/0208194.

\bibitem{strings} H. Garc\'{\i}a-Compe\'an, J.F. Pleba\'nski, M.
Przanowski and F.J. Turrubiates, ``Deformation Quantization of
Bosonic Strings'', J. Phys. A: Math. Gen. {\bf 33} (2000) 7935.

\bibitem{sw}
N.~Seiberg and E.~Witten, ``String theory and Noncommutative
Geometry,'' JHEP {\bf 9909}, 032 (1999) [arXiv:hep-th/9908142].


\bibitem{berezinbook} F.A. Berezin, {\it The Method of Second
Quantization}, (Academic Press, New York, 1966).

\bibitem{casa} R. Casalbuoni, ``The Classical Mechanics for Bose- Fermi
Systems'', Nuovo Cimmento {\bf 33} (1976) 389.

\bibitem{marinov} F.A. Berezin and M.S. Marinov, ```Particle Spin
Dynamics as the Grassmann Variant of Classical Mechanics'', Ann.
Phys. {\bf 104} (1977) 336.

\bibitem{sberezin} F.A. Berezin, {\it Introduction to Superanalysis}, D.
Reidel Publishing Company, Dordrecht (1987).

\bibitem{dewittbook} B. deWitt, {\it Supermanifolds}, Second Edition
Cambridge University Press, Cambridge (1992).

\bibitem{susyqmb} A. Lahiri, P. Kumar Roy and B. Bagchi,
``Supersymmetry in Quantum Mechanics'', Int. J. Mod. Phys. A {\bf
5} (1990) 1383.

\bibitem{bereone} F.A. Berezin, Theor. Math. Phys. (USSR) {\bf 6} (1971)
94.

\bibitem{beretwo} F.A. Berezin, Proc. Moscow Math. Soc. {\bf 17} (1967)
117.

\bibitem{schmutz} M. Schmutz, Nuovo Cimento B {\bf 25} (1975) 337.

\bibitem{marnelius} R. Marnelius, ``Half-Integer Ghost States and
Simple BRST Quantization'', Nucl. Phys. B {\bf 294} (1987) 671;
``Fermionic Quantum Mechanics and Superfields'', Int. J. Mod.
Phys. A {\bf 5} (1990) 329.

\bibitem{bordemann} M. Bordemann, ``The Deformation Quantization of
Certain Super-Poisson Brackets and BRST-Cohomology'', in {\it
Quantization, Deformations, and Symmetries}, Vol. II, Eds. G. Dito
and D. Sternheimer, Kluwer Academic Publishers, Dordrecht (2000).

\bibitem{bordemannII} M. Bordemann, H.-C. Herbig and S. Walmann,
Commun. Math. Phys. {\bf 210} (2000) 107.

\bibitem{duetsch}
  M.~Duetsch and K.~Fredenhagen,
  ``Deformation Stability of BRST-quantization,''
  AIP Conf.\ Proc.\  {\bf 453}, 324 (1998) [arXiv:hep-th/9807215].

\bibitem{zachosfermions} C. Zachos, J. Math. Phys. {\bf 41} (2000)
5129.

\bibitem{hirshfeld} A.C. Hirshfeld and P. Henselder, ``Deformation
Quantization for Systems with Fermions'', Ann. Phys. (N.Y.) {\bf
302} (2002) 59.

\bibitem{clifford} A.C. Hirshfeld and P. Henselder, ``Clifforddization,
Spin and Fermionic Star Products'', Annals Phys. {\bf 314} (2004)
75, quant-ph/0404168.

\bibitem{imelda}
  I.~Galaviz, H.~Garc\'{\i}a-Compe\'an, M.~Przanowski and F.~J.~Turrubiates,
  ``Weyl-Wigner-Moyal Formalism for Fermi Classical Systems,''
  arXiv:hep-th/0612245.


\bibitem{seiberg} N. Seiberg, ```Noncommutative Superspace, ${\cal
N}={1 \over 2}$ Supersymmetry, Field Theory and String Theory'',
JHEP {\bf 0306}, 010 (2003), [arXiv:hep-th/0305248].

\bibitem{wwmformalism} R.L. Stratonovich, {\it Sov. Phys. JETP}
{\bf 31}, 1012 (1956); A. Grossmann, {\it Commun. Math. Phys.}
{\bf 48}, 191 (1976) ; J.M. Gracia-Bond\'{\i}a, ``Hydrogen Atom in
the Phase-space Formulation of Quantum Mechanics'', Phys. Rev.  A
{\bf 30} (1984) 691; J.M. Gracia Bond\'{\i}a and J.C. Varilly,
{\it J. Phys. A: Math. Gen.} {\bf 21}, L879 (1988), {\it Ann.
Phys.} {\bf 190}, 107 (1989); J.F. Cari\~nena, J.M. Gracia
Bond\'{\i}a and J.C. Varilly, {\it J. Phys. A: Math. Gen.} {\bf
23}, 901 (1990): M. Gadella, M.A. Mart\'{\i}n, L.M. Nieto and M.A.
del Olmo, {\it J. Math. Phys.} {\bf 32}, 1182 (1991); J.F.
Pleba\'nski, M. Przanowski and J. Tosiek, {\it Acta Phys. Pol.}
{\bf B27} 1961 (1996).

\bibitem{tata} W.I. Tatarskii, {\it Usp. Fiz. Nauk} {\bf 139}, 587
(1983).

\bibitem{hillery} M. Hillery, R.F. O'Connell, M.O. Scully and E.P.
Wigner, {\it Phys. Rep.} {\bf 106}, 121 (1984).

\bibitem{wein} S. Weinberg, {\it The Quantum Theory of Fields} Vol. I,
(Cambridge University Press, Cambridge, 1995).

\bibitem{hatfield} B. Hatfield, {\it Quantum Field Theory of Point Particles and
Strings}, Addison-Wesley Publishing Company, 1992.

\bibitem{qed} I. Galaviz, H. Garc\'{\i}a-Compe\'an, M.
Przanowski and F.J. Turrubiates, ``Deformation Quantization of
Spinor Electrodynamics'', to appear (2007).

\bibitem{fronsdal}
  C.~Fronsdal and M.~Kontsevich,
  ``Quantization on Curves,''
  arXiv:math-ph/0507021.

\bibitem{martinec}
  E.~J.~Martinec and G.~W.~Moore,
  ``Noncommutative solitons on orbifolds,''
  arXiv:hep-th/0101199.

\bibitem{belhaj}
  A.~Belhaj, J.~J.~Manjarin and P.~Resco,
  ``On non-commutative orbifolds of K3 surfaces,''
  J.\ Math.\ Phys.\  {\bf 44}, 2507 (2003)
  [arXiv:hep-th/0207160].


\end{references}
\end{document}